  \documentclass[multi]{cambridge6A}
  \usepackage{natbib}

  \usepackage{rotating}
  \usepackage{floatpag}
  \usepackage{adjustbox}
  \rotfloatpagestyle{empty}

  \usepackage{graphicx}

\alphafootnotes

\begin{document}

\setcounter{chapter}{20}
  \author[Martin\,Sahl\'en]{Martin Sahl\'en\footnotemark}
\chapter{On Probability and Cosmology: Inference Beyond Data?} 
\footnotetext[1]{Department of Physics and Astronomy, Uppsala University, SE-751 20 Uppsala, Sweden}
\contributor{Martin Sahl\'en
    \affiliation{Department of Physics and Astronomy, Uppsala University, SE-751 20 Uppsala, Sweden}}
\arabicfootnotes

\textit{Slightly expanded version of contribution to the book `The Philosophy of Cosmology', Khalil Chamcham, Joseph Silk, John D. Barrow, Simon Saunders (eds.), Cambridge University Press, 2017.}

\section{Cosmological Model Inference with Finite Data} 
In physical cosmology we are faced with an empirical context of gradually diminishing returns from new observations. This is true in a fundamental sense, since the amount of information we can expect to collect through astronomical observations is finite, owing to the fact that we occupy a particular vantage point in the history and spatial extent of the Universe. Arguably, we may approach the observational limit in the foreseeable future, at least in relation to some scientific hypotheses \citep{Ellis:2014rp}. There is no guarantee that the amount and types of information we are able to collect will be sufficient to statistically test all reasonable hypotheses that may be posed. There is under-determination both in principle and in practice \citep{Zinkernagel:2011nq,Butterfield:2014hm,Ellis:2014rp}.  These circumstances are not new, indeed cosmology has had to contend with this problem throughout history. For example, \cite{1949suic.book.....W} relates the same concerns, and points back to remarks by Blaise Pascal in the 17th century: ``But if our view be arrested there let our imagination pass beyond; ... We may enlarge our conceptions beyond all imaginable space; we only produce atoms in comparison with the reality of things.'' Already with Thales, epistemological principles of uniformity and consistency have been used to structure the locally imaginable into something considered globally plausible. The primary example in contemporary cosmology is the Cosmological Principle of large-scale isotropy and homogeneity. In the following, the aim will be to apply principles of uniformity and consistency to the procedure of cosmological model inference itself.

The state of affairs described above naturally leads to a view of model inference as inference to the best explanation/model \citep[e.g.][]{Maher:1993ft,Lipton:2004zj}, since some degree of explanatory ambiguity appears unavoidable in principle. This is consistent with a Bayesian interpretation of probability which includes \textit{a priori} assumptions explicitly. As in science generally, inference in cosmology is based on statistical testing of models in light of empirical data. A large body of literature has built up in recent years discussing various aspects of these methods, with Bayesian statistics becoming a standard framework \citep{2003prth.book.....J,Hobson:2010vh,Toussaint:2011ud}. The necessary foundations of Bayesian inference will be presented in the next section. 

Turning to the current observational and theoretical status of cosmology, a fundamental understanding of the dark energy phenomenon is largely lacking. Hence we would like to collect more data. Yet all data collected so far point consistently to the simplest model of a cosmological constant, which is not well-understood in any fundamental sense. Many of the other theoretical models of dark energy are also such that they may be observationally indistinguishable from a cosmological constant \citep{Efstathiou:2008if}. Another important area is empirical tests of the inflationary paradigm, the leading explanation of the initial conditions for structure in the Universe \citep[e.g.][]{Smeenk:2014sj}. Testing such models necessitates, in principle, the specification or derivation of an \textit{a priori} probability of inflation occurring (with particular duration and other relevant properties). The \textit{measure problem} is the question of how this specification is to be made, and will be the departure point and central concern in the following sections. 

We will argue that the measure problem, and hence model inference, is ill defined due to ambiguity in the concepts of probability, global properties, and explanation, in the situation where additional empirical observations cannot add any significant new information about some relevant global property. We then turn to the question of how model inference can be be made conceptually well-defined in this context, by extending the concept of probability to general valuations (under a few basic restrictions) on partially ordered sets known as lattices. On this basis, an extended \textit{axiological Bayesianism} for model inference is then outlined. The main purpose here is to propose a well-motivated, systematic formalisation of the various model assessments routinely, but informally, performed by practising scientists.

\section{Bayesian Inference}
Inference can be performed on different levels. An important distinction is that between parameter and model inference: the first assumes that a particular model is true and derives the most likely model parameter values on the basis of observations, whereas the latter compares the relative probability that different models are true on the basis of observations. The two inferential procedures can be regarded as corresponding epistemically to description (parameter inference) and explanation (model inference) respectively. This chapter will focus on model inference, which becomes particularly troublesome in the global cosmological context. We present both cases below for completeness. For more on Bayesian inference, see e.g. \cite{2003prth.book.....J}.

\subsection{Parameter Inference}
Bayesian parameter inference is performed by computing the \textit{posterior probability}
\begin{equation}
p(\theta | D; M) = \frac{\mathcal{L}(D | \theta; M) \Pi(\theta; M)}{P(D; M)}\,,
\end{equation}
where $D$ is some collection of data, $M$ the model under consideration, and $\theta$ the model parameters. The \textit{likelihood} of the data is given by $\mathcal{L}(D | \theta; M)$ and the \textit{prior} probability distribution is $\Pi(\theta; M)$. The normalisation constant $P(D; M)$ is irrelevant for parameter inference, but central to model inference, and will be discussed next. The expression above is known as \textit{Bayes' theorem}. 

\subsection{Model Inference}
The \textit{Bayesian evidence} for a model $M$ given data $D$ can be written
\begin{equation}
\label{eq:evidence}
P(D; M) = \int \mathcal{L}(D | \theta; M) \Pi(\theta; M) d\theta\,,
\end{equation}
where symbols are defined as above. The Bayesian evidence is also called the \textit{marginalised likelihood}, reflecting the fact that it measures the average likelihood across the prior distribution, and is thus a measure of overall model goodness in light of data and pre-knowledge. It is used in inference to compare models, with a higher evidence indicating a better model. Conventional reference scales (e.g. the Jeffreys scale) exist to suggest when a difference in evidence is large enough to prefer one model over another \citep{Hobson:2010vh}.

Looking at Eq.~(\ref{eq:evidence}), the Bayesian evidence is clearly sensitive to the specification of the prior distribution. A prior is usually specified based on {\it previous} empirical knowledge from parameter estimation, or may be predicted by the theoretical model, or given by some aesthetic principle. This highlights two things: without empirical pre-knowledge, a prior is entirely based on theoretical or philosophical assumption, and a prior is also not cleanly separable from the model likelihood and can therefore to a degree be regarded as part of the model.  As increasing amounts of data is collected, the influence of the initial prior is gradually diminished through the process of \textit{Bayesian updating}, i.e. the current posterior probability becomes the (new) prior probability for a future data analysis. Through this process, the posterior eventually converges to a distribution essentially only dependent on the total numbers of and precisions of measurements. Increasingly numerous and precise measurements make the initial prior insignificant for the posterior. When data is extremely limited relative to the quantity/model of interest, this process stops short and the initial prior can then play a significant role in the evidence calculation. This will be of importance in the discussion that follows.

\section{Global Model Inference in Cosmology}
Cosmology, by its nature, seeks to describe and explain the large-scale and global properties of the Universe. There is also, by the nature of the field, a problem of finite data and underdetermination that becomes particularly poignant for measuring and explaining some global properties of the Universe. This will typically be associated with features on physical scales corresponding to the size of the observable Universe or larger, or features in the very early Universe. On the one hand, there is an epistemological question of knowledge based on one observation (i.e. the one realisation of a universe we can observe): how accurate/representative is our measurement? On the other hand, there is an ontological question of whether a property is `truly' global: if not, how might it co-depend on other properties, with possible implications for the evaluation of probabilities and inference? We shall therefore distinguish epistemically and ontically global properties in the following. In general, a global property will be defined here as some feature of the Universe which remains constant across the relevant domain (e.g. observable Universe, totality of existence). 

A conventional approach to global properties is to treat separate regions of the Universe as effectively separate universes, such that sampling across regions in the Universe corresponds to sampling across different realisations of a universe. While this approach is useful for understanding the statistics of our Universe on scales smaller than the observable Universe, when approaching this scale the uncertainty becomes increasingly large and eventually dominates. This uncertainty is commonly called \textit{cosmic variance} \cite[see e.g.][]{2009pdp..book.....L}. \textbf{We will explicitly only be concerned with the case when this cosmic variance cannot be further reduced by additional empirical observations to any significant degree, for some global property of interest.}

A case in point of particular contemporary relevance concerns the initial conditions of the Universe -- what is the statistical distribution of such initial conditions? This is often described as `the probability of the Universe', or `the probability of inflation' since the inflationary paradigm is the leading explanation for producing a large, geometrically flat universe and its initial density fluctuations. More formally, it is known as the \textit{measure problem}: what is the probability measure on the space of possible universes (known as \textit{multiverse})? The measure problem is important, because parameter inference might non-negligibly depend on this measure, and model inference should non-negligibly depend on this measure. Meaningfully performing inference at this level of global properties therefore depends on finding some resolution for how to approach the measure problem. In recent years, this has led to intense debate on the scientific status of inflation theory, string theory, and other multiverse proposals \citep{Carr:2007fj, Steinhardt:2011ik, Dawid:2013ew, Kragh:2014yd, Smeenk:2014sj, Ellis:2014ti,Dawid:2015rc}.

This is not the place for addressing the range of approaches to this problem in the literature. Proposals commonly rely on some relative spatial volume, aesthetic/theoretical principle (e.g. Jeffreys prior, maximum entropy), or dynamical principle (e.g. energy conservation, Liouville measure). The reader is referred to \cite{Carr:2007fj,Smeenk:2014sj} and references therein for more details.

\section{The Measure Problem: A Critical Analysis}
\subsection{Preliminaries}
It is helpful to recognise that the measure problem is a sub-problem, arising in a particular context, related to the broader question of how to perform model inference in relation to global properties of the Universe. It arises as a problem from the desire to provide explanation for some certain global properties of the Universe, and so depends on a view of what requires explanation and what provides suitable explanation. In pursuing statistical explanation, the problem naturally presents itself through the application of conventional Bayesian statistical inference as we have seen above, and particularly in the calculation of Bayesian evidence, where the assignment of a prior probability distribution for parameter values is essential.  The model and/or prior will also explicitly or implicitly describe how different global properties co-depend, and more generally prescribe some particular structure for the unobserved ensemble of universes (hence, the multiverse). This ensemble may or may not correspond to a physically real structure. 

In addressing the measure problem, one might therefore explore the implications of varying the conditions, assumptions, and approaches described above. To what extent is the measure problem a product thereof? We will considering this question in the following. The analysis echoes issues raised in e.g. Ellis' and Aguirre's contributions in \cite{Carr:2007fj}; \cite{Ellis:2014rp}; and \cite{Smeenk:2014sj}, while providing a new context and synthesis. In the following, Kolmogorov probability will be contrasted with Bayesian probability for illustration and motivation. We note that other foundations and definitions of probability, which we do not consider, also exist (e.g., de Finetti's approach) -- see \cite{2003prth.book.....J}.

\subsection{Analysis}
\subsubsection{Structure of Global Properties}
Statistical analysis for global properties in cosmology typically relies on certain unspoken assumptions. For example, it is commonly assumed that the constants of Nature are statistically independent, or that a multiverse can be meaningfully described by slight variations of the laws of physics as we know them -- for example as in \cite{Tegmark:2006oq}. For many practical purposes, these assumptions are reasonable or irrelevant. However, in some cosmological contexts, especially in relation to global properties and questions of typicality/fine-tuning, such assumptions can impact on the conclusions drawn from observations. For example, it has been argued that fine-tuning arguments rely on untestable theoretical assumptions about the structure of the space of possible universes \citep{Ellis:2014rp}. 

A distinction was made in the preceding Section between epistemically global and ontically global properties of the Universe. A central point is that it is impossible to make this distinction purely observationally:  the set of ontically global properties will intersect the set of epistemically global properties, but which properties belong to this intersection set cannot be determined observationally. Hence, it is possible that some ontically global properties remain unknown, and that some epistemically global properties are not ontically global. This implies that in general, global properties will be subject to an uncertainty associated with these sources of epistemic indeterminacy. 

In consequence, when seeking to determine and explain some certain global properties through analysis of observational data, the possibility that the values of these global properties could depend on some other global properties -- known or unknown -- cannot be excluded empirically. One example of this possibility concerns the constants of Nature, whose values may be interdependent (as also predicted by some theories). Another example is the distinction between physical (global) law and initial conditions of the Universe: in what sense are these concepts different? They are both epistemically global properties, and from an epistemological point of view clearly interdependent to some extent (e.g. `observable history = initial conditions + evolution laws'). Yet their epistemic status is often considered to be categorically different, on the basis of extrapolated theoretical structure. These issues are discussed further in e.g.~\cite{Ellis:2014rp, Smeenk:2014sj}.

To explore these ideas further, let us consider a cosmological model described by some global parameters $\theta_{\rm p}$ (e.g. constants of Nature or initial conditions). This model also contains a specification of physical laws, which normally are considered fixed in mathematical form. For the sake of argument, let us now assume that deviations from these laws can be meaningfully described by some additional set of parameters $\delta \theta_{\rm l}$. Such a $\delta \theta_{\rm l}$ will give rise to a shift $\delta \mathcal{L}(\theta_{\rm p}, \delta \theta_{\rm l})$ in the data likelihood for our assumed exhaustive observations, relative to the same likelihood assuming standard physical laws. 

The function $\delta \mathcal{L}$ will in principle depend on the relationship between $\theta_{\rm p}$ and $\delta \theta_{\rm l}$ in some more general model picture (we have no reason \textit{a priori} to exclude such a more general picture). The shift $\delta \theta_{\rm l}$ should also affect the prior $\Pi(\theta_{\rm p})$. While this may not be explicitly stated, a parameter prior for $\theta_{\rm p}$ is generically specified conditional on the assumption of certain physical laws. For the case of global parameters, the distinction between parameters and laws becomes blurred, as they are both global properties: they may in fact be co-dependent in an extended or more general picture. However, the correlation matrix (or other dependence) between $\theta_{\rm p}$ and $\delta \theta_{\rm l}$ cannot be independently determined from observations, since only one observable realisation of parameter values is available (i.e. our one observable Universe). Hence, the shift $\delta \theta_{\rm l}$ should in general induce a shift $\Pi \rightarrow \Pi + \delta \Pi$ in the assumed prior, due to the empirically allowed and theoretically plausible dependencies between $\theta_{\rm p}$ and $\delta \theta_{\rm l}$. On this basis, there will in general be some function $\delta \Pi$ that should be included for probabilistic completeness. But this function is essentially unconstrained since it cannot be independently verified or falsified. Hence, we are in principle always free to renormalise the Bayesian evidence by an arbitrary (non-negative) amount without violating any empirical constraints. Model inference in the conventional sense therefore becomes ill defined/meaningless in this context. 

The problem here is that while we know that there in general should be co-dependencies/correlations between laws/parameters, we are unable to account for them. This means that Kolmogorov's third axiom (the measure evaluated on the `whole set' equals the sum of the measures on the disjoint subsets) is generically violated due to unaccounted-for correlations between known and unknown global properties \citep[see][for details on Kolmogorov's probability theory]{kolmogorov33,2003prth.book.....J}. The axiom could be regarded as unphysical in this context. This may also lead to problems satisfying Kolmogorov's first (non-negative probabilities) and second axiom (unitarity) for the prior. While Bayesian probability based on Cox's theorem \citep{1946AmJPh..14....1C, cox61} does not require Kolmogorov's third axiom (discussed further in the following sub-section), potential problems equally arise in relation to negative probabilities and non-unitarity. Therefore, we find that probability in this context is better thought of as \textit{quasi-probability}, occurring also e.g. in the phase space formulation of quantum mechanics \citep[][]{1932PhRv...40..749W}. 
In the quantum mechanical case, quasi-probability arises due to Heisenberg's uncertainty principle. The cosmological case can be regarded as due to an effective \textit{cosmic uncertainty principle}, arising from a finite speed of light and the spatio-temporal localisation of an observer, which fundamentally limit the observer's knowledge \citep{1960Natur.186.1035M}.

\subsubsection{Foundations for Inference}
Statistical analysis of empirical measurements is the paradigm within which inference usually takes place in science, including cosmology. It is therefore important for us to consider the foundations of probability, as applicable to this process. A central distinction as regards probability is that between \textit{physical probability} and \textit{inductive probability}. The first is some putative ontic probability, the second corresponds to the epistemological evaluation of empirical data. \cite{Albrecht:2014vs} claim that there is no physically verified classical theory of probability, and that physical probability rather appears to be a fundamentally quantum phenomenon. They argue that this undermines the validity of certain questions/statements based on non-quantum probability and make them ill defined, and that a `quantum-consistent' approach appears able to provide a resolution of the measure problem. The precise relationship between physical and inductive probability is a long-standing topic of debate, which we will not go into great detail on here \citep[for a review, see][]{2003prth.book.....J}. The essential point for our discussion is the possible distinction between the two, and the idea that inductive probability can be calibrated to physical probability through repeated observations, e.g. in a frequency-of-outcome specification, or a Bayesian posterior updating process. In that way, credence can be calibrated to empirical evidence. 

This procedure fails when considering the Universe as a whole, for which only one observable realisation exists. A conventional inductive approach to probability therefore becomes inadequate (unless one assumes that the observable Universe defines the totality of existence, but this is a rather absurd and arbitrary notion which inflationary theory also contradicts). This can also be regarded as a failure to satisfy the second axiom in Kolmogorov's definition of probability: that there are no elementary events outside the sample space \citep{kolmogorov33,2003prth.book.....J}. Without a well-defined empirical calibration of sample space and evidence, one is at risk of circular reasoning where inductive probability is simply calibrated to itself (and hence, the \textit{a priori} assumptions made in the analysis). Related situations in cosmology have been discussed in the context of the inductive disjunctive fallacy by \cite{Norton:2010ng}. 

Bayesian statistics has become the standard approach in cosmology, due to the limitations of `frequentist' methods when data is scarce in relation to the tested hypotheses (the whole-Universe and Multiverse cases being the extreme end of the spectrum). The formalism of Bayesian statistics can be motivated from rather general assumptions. For example, Cox's theorem shows that Bayesian statistics is the unique generalisation of Boolean logic in the presence of uncertainty \citep{1946AmJPh..14....1C,cox61}. This provides a logical foundation for Bayesian statistics. In recent years, it has further been shown that the Bayesian statistical formalism follows from even more general assumptions \citep[the lattice probability construction, see][]{skilling,axioms1010038}, which we will return to in Section~\ref{sec:lattice}.

There is a difference between the definition of probability based on Kolmogorov's three axioms \citep{kolmogorov33,2003prth.book.....J}, and Bayesian probability based on Cox's theorem \citep{1946AmJPh..14....1C, cox61, VanHorn20033}. Both constructions are measure-theoretic in nature, but Kolmogorov probability places a more restrictive requirement on valid measures. In Bayesian probability, measures are finitely additive, whereas Kolmogorov measures are countably additive (a subset of finitely additive measures). This means that in Bayesian probability (based on Cox's theorem), in principle the measure evaluated on the full sample space need not equal the sum of the measures of all disjoint subsets of the sample space. This can be understood to mean that integrated regions under such a measure do not represent probabilities of mutually exclusive states. In the preceding subsection, we discussed this possibility in the context of unaccounted-for correlations in the structure of global properties. In the Bayesian statistical set-up, this is thus not a problem in principle, although problematic negative probabilities and non-unitarity could also occur in this case. 

We can thus see certain benefits with a modern Bayesian statistical framework, relative to the Kolmogorov definition of probability, even though some issues also present themselves. This leaves, at least, an overall indeterminacy in `total probability' (through $\delta \Pi$). More broadly, even if Cox's theorem or the lattice probability construction provide mutually consistent logical foundations for Bayesian statistics, it remains an open question what the status of such logical foundations is. Is there a unique logical foundation? What is its relation to the physical Universe - is it a physical property? \citep[cf.][]{Putnam1969, dummett76} What is its relation to the Multiverse - is there a unified logical foundation for the Multiverse? Is \textit{cosmic logic} and \textit{multiversal inference} well-founded? \citep[e.g.][Ch. 4]{2016JCAP...02..006V,Sahlen:thesis}

\subsubsection{Modes of Explanation}
In addition to the above, a fundamental question is which phenomena, or findings otherwise, that are thought to require explanation (on the basis of current theoretical understanding), and what provides explanation. In the case of the measure problem, it is often considered that the initial conditions of the Universe appear to have been very special, i.e. in some sense very unlikely, and therefore need to be explained \citep{Ellis:2014rp, Smeenk:2014sj}. This proposition clearly rests on some \textit{a priori} notion of probable universes, often based on extrapolations of known physics (e.g. fixed global laws). 

The main mode of explanation in cosmology is based on statistical evaluation of observational data. This is usually done using the formalism of Bayesian statistics. The conventional approaches for providing a solution to the measure problem are also statistical/probabilistic in nature, and can be regarded as picking some `target' probability regime that is to be reached for the posterior probability, for a proposed measure to be considered explanatory (another alternative is some strong theoretical/structural motivation). 
There are broadly speaking two modes of explanation in this vein: ``chance'' and ``necessity''. For example, the measured values of physical constants are to be highly probable (fine-tuning/anthropics), average (principle of mediocrity), or perhaps necessarily realised at least `somewhere' (string-theory landscape). However, in view of the discussion in the preceding subsections, it appears impossible to independently establish such probabilities, and hence the notions of and distinctions between chance and necessity become blurred. Statistical explanation therefore suffers from the same problems as detailed in the preceding sub-sections, with the risk for circular confirmatory reasoning that is not actually explanatory. This critique echoes common objections to the epistemic theories of justification called \textit{foundationalism} and \textit{coherentism}. Epistemic justification based on coherentism can provide support for almost anything through circularity, and foundationalism can become arbitrary through the assertion of otherwise unjustified foundational beliefs. Neither of these two epistemological approaches appear able to provide satisfactory epistemic justification in response to the ambiguity in the effective Bayesian prior that we are discussing.

In terms of model structure, the typical form of explanation takes the shape of evolutionary, universal laws combined with some set of initial conditions at the `earliest' time. Some additional aesthetic/structural criteria such as simplicity, symmetry, and conserved quantities may also be implicitly invoked. These are usually introduced as part of the theoretical modelling, rather than as intrinsically integrated with the inferential procedure. Therefore, possible co-dependencies with other explanatory criteria (which may be thought of as a type of global properties!) are usually not considered or explored.

In conclusion, statistical explanation is ill defined in the context of the measure problem in Bayesian statistics, and the relation to other possible explanatory principles typically neglected. How to interpret a Bayesian evidence value, or differences between such values for different models, is therefore here not clear. 

\subsection{A Synthesis Proposal}	
We thus find that the measure problem is ill defined, in principle, in conventional Bayesian statistical inference in the measure problem context. This is due to a compound ambiguity in the definition of probability, prior specification, and evidence interpretation - based on the observations above that the concepts of
\begin{itemize}
\item laws and global parameters / initial conditions
\item probability
\item explanation
\end{itemize}
are ambiguous when considering a global, whole-Universe (or Multiverse) context. Hence, measure problem solution proposals in the Bayesian statistical context ultimately are subjective statements about how `strange' or `reasonable' one considers the observable Universe to be from some particular point of view. This suggests that conventional Bayesian model inference concerning global properties in the circumstances we consider is non-explanatory, and at best only self-confirmatory. Additional information is needed to provide meaningful epistemic justification. 

The ambiguities listed above can be regarded, in the language of conventional Bayesian inference, as each introducing some effective renormalisation in the expression for Bayesian evidence. Can these ambiguities be accommodated in such a way that inference becomes conceptually well-defined? There appear to be three possible, consistent, responses to this: 
\begin{enumerate}
\item Declare model inference meaningless in this context and in the spirit of Wittgenstein stay silent about it.
\item Introduce additional philosophical/explanatory principles that effectively restrict the prior (e.g. an anthropic principle, a dynamical or aesthetic principle on model space). 
\item Introduce novel empirical/epistemic domains (e.g. mathematical-structure space of accepted scientific theory, possibility space of accepted scientific theory).  
\end{enumerate}
This author argues that the natural and consistent approach involves a combination of 2. and 3. To pursue this, we need to explicitly address the ambiguous concepts and the nature of the ambiguity. The lattice conception of probability (reviewed in Section~\ref{sec:lattice}), including the measure-theoretic Bayesian probability as a special case, provides a way to do this. It allows the generalisation of Bayesian probability and evidence to include more general valuation functions which can encompass the ambiguity in model structure, probability and explanation in a consistent way. It also demonstrates that combining valuation functions is done uniquely by multiplication, suggesting that the identified ambiguities can be decomposed in such a way. 

In conclusion, the measure problem has an irreducible axiological component. While there may be some hope of falsifying certain extreme measures, it should however be clear that any successful resolution would need to address this axiological component. The proposed construction constitutes a natural extension of the Bayesian statistical framework, and can be motivated and naturally implemented starting from the notions of partially ordered sets and measure theory. It aims to explicitly include valuations which are conventionally left implicit, to provide a conceptually and theoretically consistent framework. The workings of such an \textit{axiological Bayesianism} will now be presented.

\section{Axiological Bayesianism}

\subsection{Lattice Probability}
\label{sec:lattice}
Kevin H. Knuth and John Skilling have developed a novel approach to the foundations of Bayesian probability, based on partially-ordered sets and associated valuations \citep{axioms1010038}. It generalises Kolmogorov's and Cox's earlier work. An overview is given in \cite{skilling}, but the main features will be outlined here. The construction starts off from a general set of possibilities, for example a set of different models or parameter values, but where our ultimate purpose is to quantitatively constrain to a sub-set of inferred preferable possibilities. The set is built up by a `null' element, a set of basis elements (e.g. \{Model A, Model B\}), and the set of all combinations of logical disjunctions (`OR') between the basis elements (e.g. \{Model A-OR-Model B\}). 

On this set, \textit{partial ordering} is defined, denoted here by `$<$'. For elements $x$ and $y$, we have that $x<y$ means that $y$ includes $x$.
The ordering is required to be transitive, i.e. 
\begin{equation}
x<y \mbox{ and } y<z \Longrightarrow x<z \,.
\end{equation}
The concept of \textit{least upper bound} is introduced separately. The least upper bound to $x$ and $y$, if it exists, is the least element at or including both $x$ and $y$. We denote it by $x \lor y$. The \textit{greatest lower bound} of $x$ and $y$ is defined analogously, and denoted $x \land y$. 

A \textit{lattice} is a partially-ordered set with a well-defined least upper bound, reflecting the idea that the ordering induces a structure on the set. It also obeys (among other conventional axioms) associativity, $(x\,\lor\,y)\,\lor\,z = x\,\lor\,(y\,\lor\,z)$. This property is central to the probability construction based on valuations that now follows. 

On the lattice, a function prescribing a quantitative valuation to each lattice element is then introduced. The purpose of this valuation is to rank elements. Requiring that such a valuation respects the ordering and the lattice structure, it can be shown that any valuation $m$ must satisfy 
\begin{equation}
m(x \lor y) = m(x) + m(y)\,,
\end{equation}
without loss of generality. This is essentially what defines a mathematical \textit{measure}. From this it also follows that the valuation of general lattice elements (constructed via use of `$\lor$') can be built up by addition from valuation prescriptions for the basis elements.

One can also consider a direct product `$\times$' of lattices, which under similar assumptions on $m$ as above leads to the requirement
\begin{equation}
m(x \times y) = m(x)m(y)
\end{equation}
on the valuation $m$. Combinations are thus always multiplicative. This will be of particular importance in the following. 

Turning to the question of how to define probability, it can be shown that under preservation of lattice structure, associativity, and unitarity, conventional Bayesian probability calculus follows, with a probability $p$ defined by
\begin{equation}
p(x | t) \equiv \frac{m(x \land t)}{m(t)}\,,
\end{equation}
where $t$ is some lattice context that one considers \textit{a priori}. This expression generalises the conventional probability concept and calculus to any valuation concordant with ordering, lattice structure and associativity. It provides therefore a basis for a generalised inference procedure. A prescription for how to reason rationally also within the ambiguous context we consider.

\subsection{Gevidence: Generalisation of Bayesian Evidence}
The concept of explanation is intrinsically tied to the concept of probability in the context of statistical explanation. In generalising the concept of probability to general valuations (which, as shown above, must also be mathematical measures in the Knuth--Skilling construction), statistical explanation can therefore in that process also be generalised. Such a generalisation will involve an evaluation of how well a model corresponds to some set of explanatory principles encoded in valuations (e.g. predicting empirical observations, satisfying aesthetic criteria, etc.). We therefore turn to the question of how Bayesian evidence can be generalised on the basis of the lattice probability construction, to provide a resolution of the conceptual problems associated with the measure problem. A key question for implementing a generalisation of Bayesian evidence for model inference, is how to combine several valuations corresponding to different explanatory criteria and empirical/epistemic domains, into a compound `net' valuation. The preceding subsection gives the unique answer: by multiplication. Given a set of valuations/lattice representations, there is thus a unique way to define a probability based on these, through multiplication.

In analogy with the way in which different physical measurements can be combined to form joint likelihoods, other explanatory criteria can be combined in different ways using the multiplicative prescription. We therefore define a generalised Bayesian evidence -- let us call it \textit{gevidence} for short -- by 
\begin{equation}
P(a,b,c,...; M) \equiv \int p(a | \theta; M) p(b | \theta; M) p(c | \theta; M) ... \Pi(\theta; M) d\theta \,,
\end{equation}
where the letters $a$, $b$, $c$, ... refer to different valuation prescriptions corresponding to explanatory criteria. It also useful to define the log-quantity
\begin{equation}
L_{abc...} \equiv \log P(a,b,c,...; M) \,,
\end{equation}
which is more useful when performing model comparison, since the log-quantities add/subtract between models. 

While any valuation measure in itself cannot be `proven', just like for a parameter prior it can be founded on theoretical principle and experience. The novel measures are fundamentally no different from conventionally used priors on model parameter space - they simply generalise to higher orders of model characteristics. The proposed construction provides a prescription for how to carry out model comparison on the basis of such measures. It is not primarily intended as a tool to exclude models, but rather a means of maintaining a principled and systematic approach to comparing models, given certain assumptions. Note however that it is possible to perform inference on the explanatory criteria through re-conditionalising, as discussed in Section~\ref{sec:implappl}. 

\subsection{Evidence, Elegance, Beneficence}
  
While the proposed construction does not remove epistemic ambiguity, it provides a rational, natural and well-defined framework for examining this ambiguity in a systematic, rational and explicit way to determine relative model fitness. A rough way to categorise the possibilities is on the basis of empirical/epistemic domain. A practical way to classify models is then by theoretical/mathematical structure and physical possibility space. This can be translated to correspond to aesthetic and ethical principles. While this terminology may appear unorthodox, it emphasises the axiological element which is present regardless, but does not therefore imply the presence of any aesthetic or moral agent. Model inference thus divides into interlinking empirical, aesthetic and ethical comparisons. This classification may also be useful in that it reflects how scientists intuitively tend to approach informal model assessment. To structure the problem, let us therefore represent the gevidence in the specific form
\begin{equation}
P(D, A, E; M) = \int \mathcal{L}(D | \theta; M) p(A | \theta; M) p(E | \theta; M) \Pi(\theta; M) d\theta \,,
\end{equation}
where $D$ denotes ``data'', $A$ denotes ``aesthetics'', and $E$ denotes ``ethics''. We thus represent models on a direct product set of empirical observables, model structure, and model possibility space. Each of these probabilities may in themselves be subdivided by multiplication into any number of component valuations. 

We may also form the partial gevidences $P(A, D; M)$, $P(A, E; M)$, and $P(D, E; M)$, which provide additional information about the ways in which the different explanatory criteria corroborate or contradict each other. We shall refer to the individual gevidences as \textit{evidence} [$P(D; M)$], \textit{elegance} [$P(A; M)$], and \textit{beneficence} [$P(E; M)$]. The  log-quantities $L_{ADE}$, $L_{AD}$, $L_{AE}$, $L_{DE}$, $L_D$, $L_A$, and $L_E$ will also be of interest for model comparison. 

This quantification offers a formalisation of model comparison and explanation across the categories, and hence also of problem formulation: just as an unexpectedly rare state in model phase space (fine-tuning) may prompt explanation, e.g. an unexpectedly un-aesthetic/aesthetic model structure may prompt explanation. 

\subsection{Implementation and Application}
\label{sec:implappl}
Let us now turn to how, in practice, the type of framework proposed could be implemented and used. A few essential elements are needed:
\begin{itemize}
\item Basis elements for lattice (representation of models);
\item Aesthetic measure/s on model-structure space;
\item Ethical measure/s on model possibility space;
\item Computational capability to evaluate on model-structure space and possibility space.
\end{itemize}
Out of these, the first three appear straightforward to achieve. Some aesthetic measures of e.g. simplicity are naturally incorporated in the conventional Bayesian statistical framework (Ockham's razor), others may conventionally be used more informally. The proposed framework formalises model structure considerations beyond `parameter shaving' with Ockham's razor. A list of some possible aesthetic criteria \citep[after][]{Ellis:2014rp} are shown in Table~\ref{tabmodcrit}. \begin{table}[htdp]
\caption[Example aesthetic model comparison criteria]{Example aesthetic model comparison criteria, after \cite{Ellis:2014rp}.}
\label{tabmodcrit}
\begin{center}
\begin{tabular}{|l|l|}
& \underline{Satisfactory structure}\\
& (a) internal consistency, 
(b) simplicity (Ockham's razor), \\
& (c) `beauty' or `elegance'; \\
\\
& \underline{Intrinsic explanatory power}\\
 & (a) logical tightness, \\
 & (b) scope of the theory - unifying otherwise separate phenomena; \\ 
 \\
& \underline{Extrinsic explanatory power}\\
 & (a) connectedness to the rest of science,\\
 & (b) extendability - a basis for further development.\\
\end{tabular}
\end{center}
\label{default}
\end{table}%
As an example, ``connectedness to the rest of science'' might be quantified on the basis of the different physical constants that appear in a model. \cite{Ellis:2014qt} argue that Higgs inflation could be regarded as a preferred inflation model on such grounds.  Ethical measures are not commonly discussed, although scientists (as everyone) will at some level be influenced by such considerations, for example due to their philosophical position along the axis from materialism to idealism (ethics defined on exclusively materialistic or idealistic grounds can clearly differ significantly). Explicit consideration of ethics in relation to cosmology is given by \cite{Knobe:2006qi}, who discuss the ethical implications of inflationary cosmology, and in \cite{murphy96} where it is argued that scientific cosmology points toward a kenotic ethic. Computational capability on model-structure space should be reasonably adequate with today's technology. However, computations of/on possibility spaces may present serious challenges especially for complex theories and measures. 
\begin{figure}
  \adjustbox{trim={0.24\width} {.29\height} {0.2\width} {.23\height},clip}%
  {\includegraphics[scale=0.64]{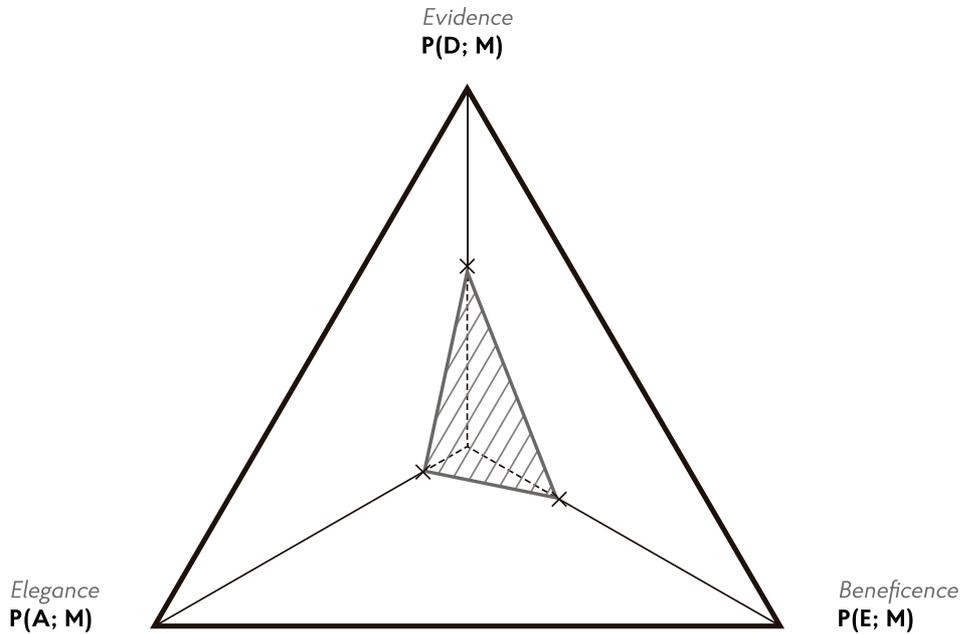}}
    \caption[The Axiological Bayesian Triangle]
      {A schematic axiological Bayesian triangle representation of model evidence $P(D; M)$, elegance $P(A; M)$, and beneficence $P(E; M)$.}
    \label{triangle}
  \end{figure}
 \begin{figure}
    \adjustbox{trim={0.28\width} {.17\height} {0.28\width} {.16\height},clip}%
  {\includegraphics[scale=0.64]{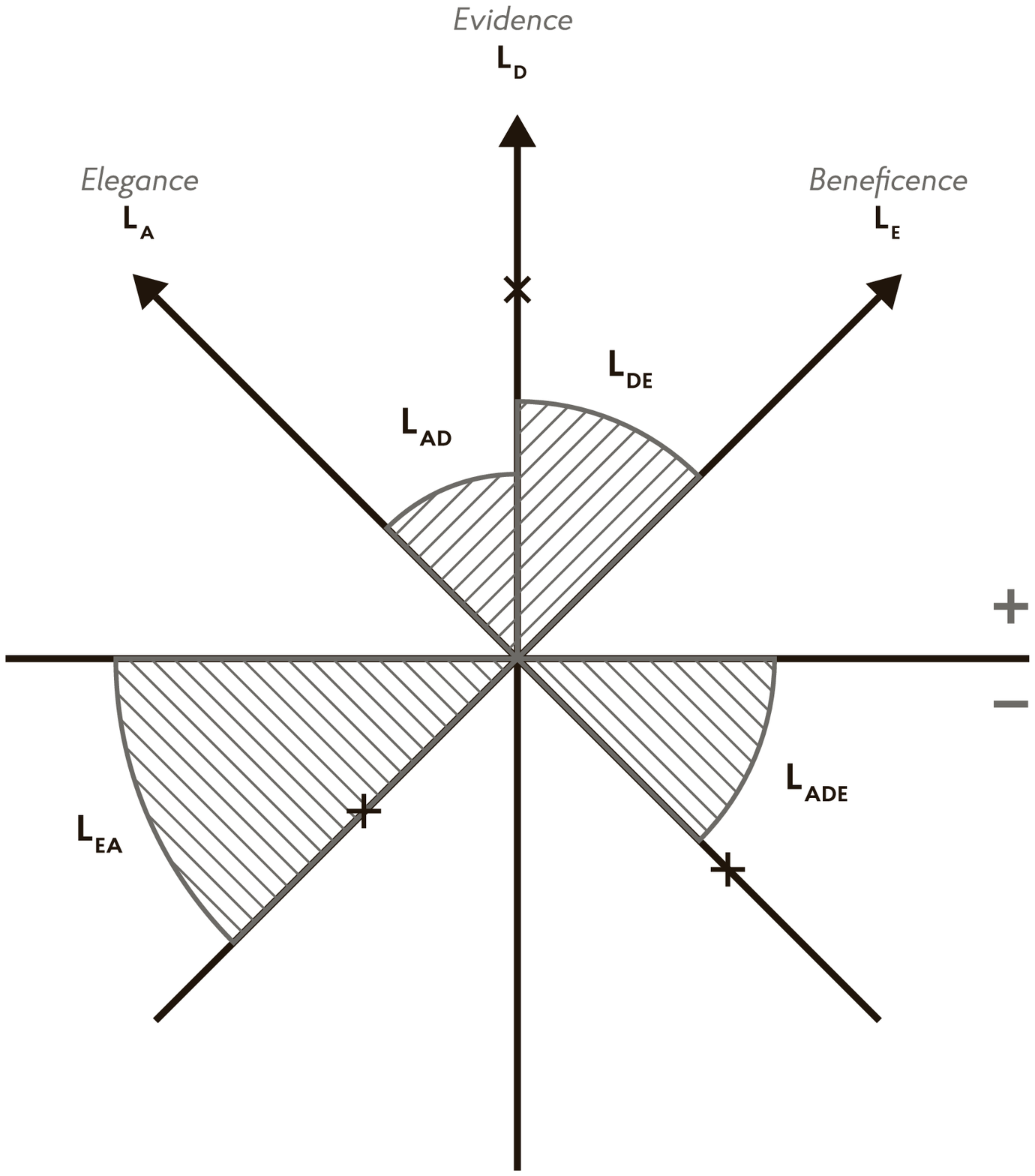}}
    \caption[The Axiological Bayesian Circle of Comparison]{A schematic axiological Bayesian circle representation of model gevidence, useful for model comparison. The implicit model conditioning has been dropped in the figure. The figure is divided into a positive and negative half-plane. There are three axes on which are crosses to indicate the values of $L_D$, $L_A$ and $L_E$. These axes and the axis separating the half-planes divide the figure into eight equal  segments. Within these are plotted shaded circle segments with areas corresponding the values of $L_{AD}$, $L_{DE}$, $L_{AE}$ and $L_{ADE}$. The shaded circle segments are placed in the segment delineated by the axes corresponding to the two criteria in question, e.g. $L_{AD}$ is placed between the axes for $L_D$ and $L_A$. In the case of $AE$, the half-plane axis forms one of the axes. $L_{ADE}$ is plotted in the remaining section. Positive values are indicated by placing the shaded circle segment in the positive half-plane, and negative values in the negative half-plane. This is reinforced by choosing the orientation of the shading lines to distinguish positive and negative values. The areas can be directly compared within figures for a given model, and between figures for different models, to indicate relative model fitness (a model difference figure can also be constructed in the same way). Recall that the larger the values are, the stronger the support is for a model.}
    \label{circle}
  \end{figure}

In practice, the framework can be used much the way models are compared in the light of different data sets both separately and jointly to test consistency and combined inferential power. Fig.~\ref{triangle} shows schematically a simple way to illustrate the evidence $P(D; M)$, elegance $P(A; M)$ and beneficence $P(E; M)$. A fuller representation is shown schematically in Fig.~\ref{circle}, where the log-gevidences are shown for all possible combinations of criteria. This figure gives a complete picture of the model gevidence, and can be directly compared between models to understand which model is best overall, or with respect to combinations of only some of the criteria. This offers possibilities to explore multi-factor explanations, i.e. part empirical, part aesthetic, part ethical. It gives a clear picture of to what extent empirical data and axiological criteria are consistent with each other.

One may also examine which aesthetic and ethical criteria are `preferred' with the help of the framework. By re-conditionalising the probability, one can compute 
e.g. $P(D | A; M) = P(D, A; M)/P(A; M)$ which quantifies the empirical support for the elegance principle $A$ (under model $M$). One could also separately study the support for particular explanatory criteria across some range of models of relevance by comparing $P(C) = \Sigma_M P(C; M)\Pi(M)$ for different criteria $C$, which thus effectively extends the empirical/epistemic domain to model structure and model possibility space.

\section{Concluding Discussion}
In this Chapter, it has been argued -- in concordance with earlier observations by \cite{Ellis:2014rp}, \cite{Smeenk:2014sj}, and others -- that 
\begin{enumerate}
\item Model inference in cosmology involves both evaluation of empirical statistical evidence and application of other interpretative principles.
\item The Bayesian statistical framework, particularly, suffers from the measure problem in relation to explanation of global properties in a whole-Universe and Multiverse context (notably, inflationary initial conditions).
\item Some interpretative principles are not in themselves empirically testable by conventional Bayesian statistical tests,
\item Bayesian statistical explanation therefore is effectively qualitative in the whole-Universe context, such that Bayesian probability becomes ambiguous quasi-probability and the measure problem ill defined.
\item It is possible to extend Bayesian statistical inference in a natural and unique way to explicitly account for non-observational explanatory principles and provide a conceptually well-defined inferential procedure.
\end{enumerate}
These considerations lead to the following conclusion. If we accept probability calculus as founded on the lattice construction, then the conventional scientific method can be regarded as a special case of a more general part-subjective, \textit{but uniquely rational}, framework for reasoning we have termed ``axiological Bayesianism''. This framework generalises Bayesian statistics to define a more general version of Bayesian evidence for model inference. 
We have called this ``gevidence'' and divided it into three main sub-components: evidence, elegance, and beneficence. This enables the inclusion of probabilities based on valuation measures on model structure and possibility space, that combine in a unique way. The framework appears to have overlap with Dawid's concept of \textit{non-empirical theory evaluation} \citep{Dawid:2013ew}, and to lend itself to the epistemic theory of justification called \textit{foundherentism} \citep{HaackEvIn}, a synthesis of foundationalism and coherentism. The framework can be further justified by appealing to epistemological principles of uniformity/unity and consistency/coherence to be extended to new domains, i.e. model inference and comparison. 

Potential problems with the proposed inference approach arise if the rules of probability are themselves global empirical properties of the Universe, just like a physical law, since such a `probability law' could be different in other parts of a multiverse. The quasi-probability nature of Bayesian statistics in our analysis, and the lattice construction foundation, suggests that the framework may be extended to consider alternative logical foundations for probability, e.g. \textit{quantum logic} \citep{svozil98}, \textit{intuitionistic logic} \citep{weatherson2003}, \textit{fuzzy logic} \citep{zadeh1978fuzzy}. It remains to be seen how the framework of axiological Bayesianism might be developed and applied in practice. The details of model representation, associated product-space construction and measures, as well as computational techniques, need to be worked out. A reference scale for gevidence differences would be desirable \citep[perhaps based on the concept of \textit{information}, see][]{skilling}. 

When there are very limited data, it is inevitable in principle that some type of hermeneutic process comes into play when engaging in inference. We must then either accept additional explanatory criteria (i.e. not based on data likelihood) as valid `scientific method', or appeal to some additional principle that invalidates such criteria. One possible such principle might be that the type of subjectivity inherent in axiological Bayesianism is outside the realms of science, and hence the framework is to be rejected. This is a perfectly valid position. However, since it was shown above that subjective ambiguity is also present implicitly in the conventional Bayesian framework, this principle also excludes making statements about the measure problem using that same framework. One should then choose to stay silent on the matter, to remain consistent. Hence, the proposed framework appears to be an in principle necessary, conceptually consistent, and theoretically natural (though not necessarily unique) generalisation of the Bayesian statistical framework for addressing the measure problem and similar questions, for those who wish not to stay silent.

\section*{Acknowledgments}
Thanks are due to Jeremy Butterfield, Khalil Chamcham, Daniel Darg, George Ellis, Hans Halvorson, Andrew Liddle, Laura Mersini-Houghton, Brian Pitts, Joseph Silk, David Sloan, Chris Smeenk, and Henrik Zinkernagel, for useful conversations in this area. 

\bibliographystyle{cambridgeauthordate}

\begin{thebibliography}{40}
\expandafter\ifx\csname natexlab\endcsname\relax\def\natexlab#1{#1}\fi
\expandafter\ifx\csname selectlanguage\endcsname\relax
  \def\selectlanguage#1{\relax}\fi

\bibitem[\protect\citename{Albrecht and Phillips, }2014]{Albrecht:2014vs}
Albrecht, A, and Phillips, D. 2014.
\newblock Origin of probabilities and their application to the multiverse.
\newblock {\em Physical Review D}, {\bf 90}(12).

\bibitem[\protect\citename{Butterfield, }2014]{Butterfield:2014hm}
Butterfield, J. 2014.
\newblock On under-determination in cosmology.
\newblock {\em Studies in History and Philosophy of Modern Physics}, {\bf 46},
  57--69.

\bibitem[\protect\citename{Carr, }2007]{Carr:2007fj}
Carr, B (ed). 2007.
\newblock {\em Universe or multiverse?}
\newblock Cambridge, UK: Cambridge University Press.

\bibitem[\protect\citename{{Cox}, }1946]{1946AmJPh..14....1C}
{Cox}, R.~T. 1946.
\newblock {Probability, Frequency and Reasonable Expectation}.
\newblock {\em American J. of Physics}, {\bf 14}, 1--13.

\bibitem[\protect\citename{{Cox}, }1961]{cox61}
{Cox}, R.~T. 1961.
\newblock {\em {The Algebra of Probable Inference}}.
\newblock Baltimore, MD, USA: Johns Hopkins University Press.

\bibitem[\protect\citename{Dawid, }2013]{Dawid:2013ew}
Dawid, R. 2013.
\newblock {\em String Theory and the Scientific Method}.
\newblock Cambridge, UK: Cambridge University Press.

\bibitem[\protect\citename{Dawid, }2015]{Dawid:2015rc}
Dawid, R. 2015.
\newblock Physics theory: `Simple' or `elegant' criteria are not valid.
\newblock {\em Nature}, {\bf 518}(7539), 303.

\bibitem[\protect\citename{Dummett, }1976]{dummett76}
Dummett, M. 1976.
\newblock Is logic empirical?
\newblock {Pages  45--68 of:} Lewis, H.~D., and Anscombe, G. E.~M. (eds), {\em
  Contemporary British philosophy : personal statements. Fourth series}.
\newblock Muirhead library of philosophy.
\newblock London: Allen and Unwin.

\bibitem[\protect\citename{Efstathiou, }2008]{Efstathiou:2008if}
Efstathiou, G. 2008.
\newblock Limitations of Bayesian evidence applied to cosmology.
\newblock {\em Monthly Notices of the Royal Astronomical Society}, {\bf
  388}(3), 1314--1320.

\bibitem[\protect\citename{Ellis, }2014]{Ellis:2014rp}
Ellis, G. F.~R. 2014.
\newblock On the philosophy of cosmology.
\newblock {\em Studies in History and Philosophy of Modern Physics}, {\bf 46},
  5--23.

\bibitem[\protect\citename{Ellis and Silk, }2014]{Ellis:2014ti}
Ellis, G. F.~R., and Silk, J. 2014.
\newblock Defend the integrity of physics.
\newblock {\em Nature}, {\bf 516}(7531), 321--323.

\bibitem[\protect\citename{Ellis and Uzan, }2014]{Ellis:2014qt}
Ellis, G. F.~R., and Uzan, J-P. 2014.
\newblock Inflation and the Higgs particle.
\newblock {\em Astronomy \& Geophysics}, {\bf 55}(1), 19--20.

\bibitem[\protect\citename{Haack, }1993]{HaackEvIn}
Haack, S. 1993.
\newblock {\em Evidence and Inquiry}.
\newblock Oxford, UK: Blackwell.

\bibitem[\protect\citename{Hobson, }2010]{Hobson:2010vh}
Hobson, M.~P. 2010.
\newblock {\em Bayesian methods in cosmology}.
\newblock Cambridge, UK: Cambridge University Press.

\bibitem[\protect\citename{{Jaynes}, }2003]{2003prth.book.....J}
{Jaynes}, E.~T. 2003.
\newblock {\em {Probability Theory}}.
\newblock Cambridge, UK: Cambridge University Press.

\bibitem[\protect\citename{Knobe {et~al.}, }2006]{Knobe:2006qi}
Knobe, J., Olum, K.~D., and Vilenkin, A. 2006.
\newblock Philosophical implications of inflationary cosmology.
\newblock {\em British J. for the Philosophy of Science}, {\bf 57}(1), 47--67.

\bibitem[\protect\citename{Knuth and Skilling, }2012]{axioms1010038}
Knuth, K.~H., and Skilling, J. 2012.
\newblock Foundations of Inference.
\newblock {\em Axioms}, {\bf 1}(1), 38.

\bibitem[\protect\citename{Kolmogorov, }1933]{kolmogorov33}
Kolmogorov, A.~N. 1933.
\newblock {\em Grundbegriffe der Wahrscheinlichkeitrechnung}.
\newblock Ergebnisse der Mathematik und Ihrer Grenzgebiete, vol. 2.
\newblock Berlin-Heidelberg: Springer.

\bibitem[\protect\citename{Kragh, }2014]{Kragh:2014yd}
Kragh, H. 2014.
\newblock Testability and epistemic shifts in modern cosmology.
\newblock {\em Studies in History and Philosophy of Modern Physics}, {\bf 46},
  48--56.

\bibitem[\protect\citename{Lipton, }2004]{Lipton:2004zj}
Lipton, P. 2004.
\newblock {\em Inference to the best explanation}. 2nd edn.
\newblock London, UK: Routledge.

\bibitem[\protect\citename{{Lyth} and {Liddle}, }2009]{2009pdp..book.....L}
{Lyth}, D.~H., and {Liddle}, A.~R. 2009.
\newblock {\em {The Primordial Density Perturbation}}.
\newblock Cambridge, UK: Cambridge University Press.

\bibitem[\protect\citename{Maher, }1993]{Maher:1993ft}
Maher, P. 1993.
\newblock {\em Betting on theories}.
\newblock Cambridge, UK: Cambridge University Press.

\bibitem[\protect\citename{{McCrea}, }1960]{1960Natur.186.1035M}
{McCrea}, W.~H. 1960.
\newblock {The Interpretation of Cosmology}.
\newblock {\em Nature}, {\bf 186}(June), 1035.

\bibitem[\protect\citename{Murphy and Ellis, }1996]{murphy96}
Murphy, N, and Ellis, G F~R. 1996.
\newblock {\em On the Moral Nature of the Universe}.
\newblock Theology \& the Sciences.
\newblock Minneapolis, MN, USA: Fortress Press.

\bibitem[\protect\citename{Norton, }2010]{Norton:2010ng}
Norton, J.~D. 2010.
\newblock Cosmic Confusions: Not Supporting versus Supporting Not.
\newblock {\em Philosophy of Science}, {\bf 77}(4), 501--523.

\bibitem[\protect\citename{Putnam, }1969]{Putnam1969}
Putnam, H. 1969.
\newblock Is Logic Empirical?
\newblock {Pages  216--241 of:} Cohen, R.~S., and Wartofsky, M.~W. (eds), {\em
  Proceedings of the Boston Colloquium for the Philosophy of Science
  1966/1968}.
\newblock Boston Studies in the Philosophy of Science, vol. 5.
\newblock Dordrecht: Springer Netherlands.

\bibitem[\protect\citename{Sahl\'en, }2008]{Sahlen:thesis}
Sahl\'en, M. 2008.
\newblock {\em Bayesian constraints on dark energy and cosmic structure}.
\newblock Ph.D. thesis, University of Sussex.

\bibitem[\protect\citename{Skilling, }2010]{skilling}
Skilling, J. 2010.
\newblock {\em Foundations and algorithms}.
\newblock Ch. 1 in `Bayesian methods in cosmology', M. P. Hobson (ed.),
  Cambridge, UK: Cambridge University Press.

\bibitem[\protect\citename{Smeenk, }2014]{Smeenk:2014sj}
Smeenk, C. 2014.
\newblock Predictability crisis in early universe cosmology.
\newblock {\em Studies in History and Philosophy of Modern Physics}, {\bf 46},
  122--133.

\bibitem[\protect\citename{Steinhardt, }2011]{Steinhardt:2011ik}
Steinhardt, P.~J. 2011.
\newblock The Inflation Debate : Is the theory at the heart of modern cosmology
  deeply flawed?
\newblock {\em Scientific American}, {\bf 304}(4), 36--43.

\bibitem[\protect\citename{Svozil, }1998]{svozil98}
Svozil, K. 1998.
\newblock {\em Quantum Logic}.
\newblock Singapore: Springer.

\bibitem[\protect\citename{Tegmark {et~al.}, }2006]{Tegmark:2006oq}
Tegmark, M., Aguirre, A., Rees, M.~J., and Wilczek, F. 2006.
\newblock Dimensionless constants, cosmology, and other dark matters.
\newblock {\em Physical Review D}, {\bf 73}(2).

\bibitem[\protect\citename{Van~Horn, }2003]{VanHorn20033}
Van~Horn, K.~S. 2003.
\newblock Constructing a logic of plausible inference: a guide to Cox's
  theorem.
\newblock {\em International J. of Approximate Reasoning}, {\bf 34}(1), 3 --
  24.

\bibitem[\protect\citename{{Vanchurin}, }2016]{2016JCAP...02..006V}
{Vanchurin}, V. 2016.
\newblock {Cosmic logic: a computational model}.
\newblock {\em J. of Cosmology and Astroparticle Physics}, {\bf 2}, 006.

\bibitem[\protect\citename{von Toussaint, }2011]{Toussaint:2011ud}
von Toussaint, U. 2011.
\newblock Bayesian inference in physics.
\newblock {\em Reviews of Modern Physics}, {\bf 83}(3), 943--999.

\bibitem[\protect\citename{Weatherson, }2003]{weatherson2003}
Weatherson, B. 2003.
\newblock From Classical to Intuitionistic Probability.
\newblock {\em Notre Dame J. of Formal Logic}, {\bf 44}(2), 111--123.

\bibitem[\protect\citename{{Whitrow}, }1949]{1949suic.book.....W}
{Whitrow}, G.~J. 1949.
\newblock {\em {The structure of the universe}}.
\newblock London, UK: Hutchinson's University Library.
\newblock Chap. IX: Cosmology and the {\it a priori}.

\bibitem[\protect\citename{{Wigner}, }1932]{1932PhRv...40..749W}
{Wigner}, E. 1932.
\newblock {On the Quantum Correction For Thermodynamic Equilibrium}.
\newblock {\em Physical Review}, {\bf 40}, 749--759.

\bibitem[\protect\citename{Zadeh, }1978]{zadeh1978fuzzy}
Zadeh, L.~A. 1978.
\newblock Fuzzy sets as a basis for a theory of possibility.
\newblock {\em Fuzzy sets and systems}, {\bf 1}(1), 3--28.

\bibitem[\protect\citename{Zinkernagel, }2011]{Zinkernagel:2011nq}
Zinkernagel, H. 2011.
\newblock Some Trends in the Philosophy of Physics.
\newblock {\em Theoria-Revista De Teoria Historia Y Fundamentos De La Ciencia},
  {\bf 26}(2), 215--241.

\end{thebibliography}

\end{document}